\documentclass[twocolumn,floatfix,superscriptaddress,a4paper,showkeys,nofootinbib,reprint]{revtex4-1}

\usepackage[colorlinks=true,linktocpage=true,linkcolor=blue,citecolor=blue,allcolors=blue]{hyperref}
\usepackage{epsfig}
\usepackage{latexsym}
\usepackage[utf8]{inputenc}

\usepackage{xspace}
\usepackage{indentfirst}
\usepackage{enumitem}
\usepackage{color}
\usepackage{float}

\usepackage{setspace}
\usepackage{lipsum}

\usepackage[dvipsnames]{xcolor}
\usepackage{amsmath}
\usepackage{amssymb}
\usepackage[english]{babel}
\usepackage{url}

\newcommand{\eq}[1]{\begin{align} #1 \end{align}}

\newcommand{\der}[2]{\frac{\partial #1}{\partial #2}}

\newcommand{\rom}[1]{\uppercase\expandafter{\romannumeral #1\relax}}

\begin{document}

\title{Machine learning based approach to fluid dynamics}

\author{Kirill~Taradiy}
\affiliation{Frankfurt Institute for Advanced Studies, Giersch Science Center,
D-60438 Frankfurt am Main, Germany}
\affiliation{Xidian-FIAS International Joint Research Center, Giersch Science Center,
D-60438 Frankfurt am Main, Germany}

\author{Kai~Zhou}
\affiliation{Frankfurt Institute for Advanced Studies, Giersch Science Center,
D-60438 Frankfurt am Main, Germany}

\author{Jan~Steinheimer}
\affiliation{Frankfurt Institute for Advanced Studies, Giersch Science Center,
D-60438 Frankfurt am Main, Germany}

\author{Roman~V.~Poberezhnyuk}
\affiliation{Bogolyubov Institute for Theoretical Physics, 03680 Kyiv, Ukraine}
\affiliation{Frankfurt Institute for Advanced Studies, Giersch Science Center,
D-60438 Frankfurt am Main, Germany}

\author{Volodymyr~Vovchenko}
\affiliation{Nuclear Science Division, Lawrence Berkeley National Laboratory, 1 Cyclotron Road, Berkeley, California 94720, USA}

\author{Horst Stoecker}
\affiliation{Frankfurt Institute for Advanced Studies, Giersch Science Center, D-60438 Frankfurt am Main, Germany}
\affiliation{Institut f\"{u}r Theoretische Physik, Goethe Universit\"{a}t Frankfurt, D-60438 Frankfurt am Main, Germany}
\affiliation{
GSI Helmholtzzentrum f\"ur Schwerionenforschung GmbH, D-64291 Darmstadt, Germany}

\begin{abstract}
We study the applicability of a Deep Neural Network (DNN) approach to simulate one-dimensional non-relativistic fluid dynamics.
Numerical fluid dynamical calculations are used to generate training data-sets corresponding to a broad range of profiles to perform supervised learning with DNN. 
The performance of the DNN approach is analyzed, with a focus on its interpolation and extrapolation capabilities.
Issues such as inference speed, the networks capacities to interpolate and extrapolate solutions with limited training samples from both initial geometries and evolution duration aspects are studied in detail. 
The optimal DNN performance is achieved when its objective is set to learn the mapping between hydro profiles after a fixed value time step,
which can then be applied successively to reach moments in time much beyond the duration contained in the training.
The DNN has an advantage over the conventional numerical methods by not being restricted by the Courant criterion, and it shows a speedup over the conventional numerical methods by at least two orders of magnitude.

\end{abstract}

\pacs{ }


\date{\today}

\maketitle

\section{Introduction}

Machine learning~(ML) techniques are being adapted in an increasingly large variety of scientific and engineering disciplines.
The field of computational fluid dynamics may be particularly suitable, with various ML applications possible.
The fluid dynamical simulations play an important role in a number of scientific and engineering spheres, especially for instance relativistic codes for high energy physics \cite{Scheid:1974zz,Stoecker:1986ci,Clare:1986qj, Rischke:1995ir,Kolb:2003dz,Petersen:2008dd,Gale:2013da} and astrophysics \cite{Oechslin:2001km,Baiotti:2004wn,Janka:2006fh,Bauswein:2013yna,Baiotti:2016qnr,Hanauske:2017oxo,Most:2018eaw}, but also including molecular biology, aerospace engineering, meteorology, and many other disciplines.
It has been shown that ML can provide valuable inputs for the interpretation and simulation of relativistic fluid dynamical simulations \cite{Pang:2016vdc,Huang:2018fzn,du:2020identifying} and in the field of high energy and nuclear physics in general \cite{Bernhard:2016tnd,Utama:2016tcl,Haake:2017dpr,Zhou:2018ill,Urban:2018tqv,Mori:2017nwj,Shanahan:2018vcv,Tanaka:2017niz}.
Given this, it may be of great benefit to improve various aspects of fluid dynamical applications by means of ML.
Such improvements may comprise speeding up the time-consuming hydrodynamic simulations, as well as other tasks like enforcing constraints on conservation laws, or classification of the numerical hydro solutions.

Perhaps the most natural task is to improve the computational speed of fluid dynamical simulations.
Typically, this procedure demands a trade-off with accuracy and stability, thus, it is crucial to keep these two aspects at an acceptable level.
In many cases involving scientific and industrial applications maintaining the required accuracy level while achieving the necessary speed-up of the simulations is not possible even when the implementation is with parallelization on Graphics Processing Units (GPU’s)~\cite{chen2020gpu}.
The main issue preventing fast direct simulations comes from the Courant convergence criterion~\cite{guenther1996partial}, which limits the maximum size of the time step in numerical simulations.
Other issues include the necessity of a detailed treatment of boundary conditions in Smooth-Particle-Hydro (SPH) based methods~\cite{monaghan2005smoothed,liu2003smoothed}, and in some cases the necessity to use the dynamic viscosity models~\cite{balescu1976equilibrium}.
Numerous efforts have been made to overcome these difficulties, typically sacrificing either the resolution or the robustness of the approach. 
Alongside the 
purely numerical approaches to solving partial differential equations of fluid dynamics,
several combined methods were also developed, 
where the numerical simulations are applied to a coarse-grained system, where the detailed resolution is achieved by data augmentation based on generative adversarial networks~\cite{um2018liquid,xie2018tempoGAN}. 
Such 
combined methods have shown prominent results in the field of computer gaming and cinema, creating realistic visualizations of fluid dynamics.~\cite{zaspel2011photorealistic,Harris:HarrisGPU}

In the present work we use deep neural network (DNN) as a fast solver of fluid dynamical equations.
This allows to forgo the computationally expensive numerical simulations that would require extensive computing resources.
The DNN approach emulates the solution of partial differential equations by learning the mathematical mapping encoded in the fluid dynamical equations.
As an exploratory study, we consider one-dimensional (1D) fluid dynamics. 
The neural network maps the initial state characterized by the fluid dynamical profiles of density, velocity and pressure to their state at a later time moment.
We cover in details the various aspects of the approach such as the determination of the optimal network structure, the optimal choice of the training set, variable vs fixed time step, and the performance of the network to interpolate and extrapolate.
Additionally, we separately incorporate a denoising autoencoder, which is found to increase the network accuracy substantially. The analogy between the mapping performed by the DNN and the mapping performed by the system of fluid dynamic equations makes the current approach rather transparent in its application. It opens the possibility to interpret the performance of the DNN-mapping in fairly simple way and thus, provides greater confidence for the possibility of its implementation in applied science.

The paper is organized as follows. Section~\ref{sec-meth} describes the DNN  implementation for 1D fluid dynamics, with a focus on the network structure and the training set choice. 
Section~\ref{sec-outside} investigates the ability of the network to 
extrapolate the solution beyond the original training set. 
Section~\ref{sec-fix-steps} explores the usage of a modified network structure with a fixed time step.
Section~\ref{sec-extend-training} studies the dependence of the performance on the content of the training set.
Section~\ref{sec:noisegate} discusses the application of a noise gate filter to improve the accuracy of the DNN output.
In Sec.~\ref{sec:performance} the achievable speed-ups in the DNN performance over the conventional methods are presented.
Summary in Section~\ref{summary} closes the article.

\section{Methodology}
\label{sec-meth}

We use a DNN to emulate the solution of fluid dynamical equations.
In the present work we study a one dimensional inviscid fluid described by Eulers fluid dynamical  equations: 
\eq{
&\der{\rho}{t}+u\der{\rho}{x}+\rho\der{u}{x}=0~,\\ 
&\der{u}{t}+u\der{u}{x}+\frac{1}{\rho}\der{p}{x}=0~,\\
&\der{\varepsilon}{t}+u\der{\varepsilon}{x}+\frac{p}{\rho}\der{u}{x}=0~.
}
Here $\rho$, $u$, and $p$ are the density, hydrodynamic velocity, and pressure, respectively. $\varepsilon$, $v$,  $x$, and $t$ are the internal specific energy, specific volume, spatial coordinate, and time.
While the Eulerian case is chosen here for simplicity, conceptually the method can also be applied in more elaborate cases.

The neural network maps the profiles of the hydrodynamic velocity, density and pressure fields from an initial time moment $t_0$ to a later time $t>t_0$:
\eq{
\rho(x,t_0),u(x,t_0),p(x,t_0)\rightarrow \rho(x,t),u(x,t),p(x,t)~.
}

The fluid dynamical equations are supplemented with a polytropic equation of state, $\ p {V^\gamma } = const$, where $p,V$ are the corresponding pressure and volume
values and $\gamma$ is the isentropic expansion factor.
The calculations are performed at a constant temperature.
Knowing  $\rho(x,t)$, $u(x,t)$ and for a fixed equation of state, it is possible to exclude $p(x,t)$ from the system of dynamical equations and without additionally supplementing the equation of state (EoS) obtain all three profiles.
However, 
supplying all three $\rho(x,t_0)$, $u(x,t_0)$, $p(x,t_0)$ quantities in the input layer 
opens 
the possibility
to crosscheck
the quality of the network performance by ensuring that the profiles of pressure, velocity and density produced by the network satisfy the EoS.
Therefore, the DNN performs a mapping of all three fields, $\rho(x,t)$, $u(x,t)$, $p(x,t)$. 
While in the present study we work with the polytropic EoS, our method can be applied to other equations of state as well, for instance van der Waals~\cite{hansen1990theory}, quantum van der Waals~\cite{Vovchenko:2015vxa}, or Sugie-Lu~\cite{adachi1986development} models.

Along with the consistency with the EoS, other constraints may be introduced to improve the performance of the DNN. 
The application of such constraints for the DNN
fluid dynamics simulations
will be studied
in a future paper. 

Here we use periodic boundary conditions for all the training samples.
Other choices of boundary conditions are also possible, for instance hard walls.
A different choice of boundary conditions will require re-training of the network.

In choosing the network structure we follow the recommendations of Ref.~\cite{muller2001introduction}, which suggest that the best informational capacity of the model is achieved when the chosen network structure is a little over-descriptive.
Thus, we are using a network with 3 hidden layers and one 10\% dropout layer.
The input layer contains 151 neurons: 50 for each of the $\rho(x,t)$, $u(x,t)$, $p(x,t)$ dependencies and one for $\delta t$ time shift parameter. The output layer contains 150 neurons:  50 for each of the $\rho(x,t+\delta t)$, $u(x,t+\delta t)$, $p(x,t+\delta t)$ dependencies. 
The detailed description of the network structure and training process is presented in Appendix~\ref{app-a}.

The structure and content of the training set appear to be amongst the most important factors influencing the quality of the performance of the present approach. 
The main challenge of the training set preparation is the particular choice of the curves included in the training set.  

The structure and content of the training set have a large influence on the performance of the DNN.
It is important that the training set covers a broad variety of different profiles of density, velocity and pressure. 
To achieve the goal we consider three profile functions characterizing the density and velocity profiles,
\eq{\label{f1}
 {f_1(x)}&={a_1}{e^{{b_1}{x^2}}},  \\\label{f2}
 {f_2(x)}&={a_2}\sin \left( {{b_2}x} \right) +{c_2}, \\\label{f3}
 {f_3(x)}&={a_3}{\left| {{x}} \right|}^{\frac{1}{{{b_3}}}}{\mathop{\rm sgn}} \left( x \right) + {c_3},
}
with randomly varied values of the coefficients ${a_i}$, ${b_i}$, ${c_i}$. 
Such a set of functions allows to cover a broad variety of possible input scenarios. 
It also resembles various general characteristics peculiar to the input curves of fluid dynamics. 
Each of the functions $f_1,f_2,f_3$ is chosen randomly to characterize density or velocity profiles\footnote{The pressure profile is evaluated from $\rho(x,t)$ via the equation of state.}
The considered ranges for the randomly
selected values of coefficients are the following:
\eq{\nonumber
{a_1}&\in \left[ { - 5,5} \right],~{b_1} \in \left[ { - 500,0} \right],~{a_2}\in \left[ {0,10} \right],~{b_2}  \in \left[ { - 15,15} \right],\\
{c_2} &\in \left[ { - \pi ,\pi } \right],
{a_3}   \in \left[ { - 10,10} \right],~{b_3} \in \left[ {1,10} \right],~{c_3} \in \left[ {0,10} \right].\label{ranges}
}
Note that negative values of $f_1,f_2,f_3$ are allowed only for the velocity profiles, but not for density.

Other choices of initial profile functions are possible. These can, for instance, be based on  systems of orthogonal functions like Hermite polynomials, Bessel functions, and the various Fourier series~\cite{rudin1976principles,andrews1999special}, depending on the geometric symmetries of the system.
We leave these choices for future studies.

The DNN was trained on a set of $2.5\times 10^5$ samples. Each sample was generated via the following procedure:
\begin{enumerate}
    \item 
    The initial profiles of density and velocity at $t = t_0$ are generated by randomly choosing one of the curve families for each of the profiles independently in Eqs.~(\ref{f1})-(\ref{f3}). The coefficients defining the curve from the chosen family are being randomly generated within the defined boundaries [Eq.~(\ref{ranges})]. For the density profiles the values of the coefficients are constrained such that to ensure the positivity of the density profile.
    \item For each pair of the initial density and velocity profiles the pressure profile is computed using the EoS. 
    \item The Euler equations are solved numerically to obtain the density, pressure and velocity profiles are at a future time moment $t = t_0 + \delta t$, where $\delta t$ at this particular case is chosen to be equal to 1.  
    \item The combination of initial states at $t_0$ and final states at $t_0+\delta t$ are organized into training pairs
    \eq{
    \{\delta t;\rho_0(x),u_0(x),p_0(x)|\rho_{\delta t}(x),u_{\delta t}(x),p_{\delta t}(x)\}
    }
    and incorporated into the training set.
\end{enumerate}

The DNN training procedure consists of $23$ training epochs, where the saturation of training and validation error is observed.
The resulting training set accuracy level, quantified by the $R^2$ metric, is about 89\%.
We explored also a simpler DNN structure consisting of only a single hidden layer, in that case the DNN performance was inferior, with the accuracy level not exceeding 83\%.
Thus, the structure with 3 hidden layers and a 10\% dropout layer was found to be reasonable for the task at hand.

\begin{figure*}
\includegraphics[width=1\textwidth]{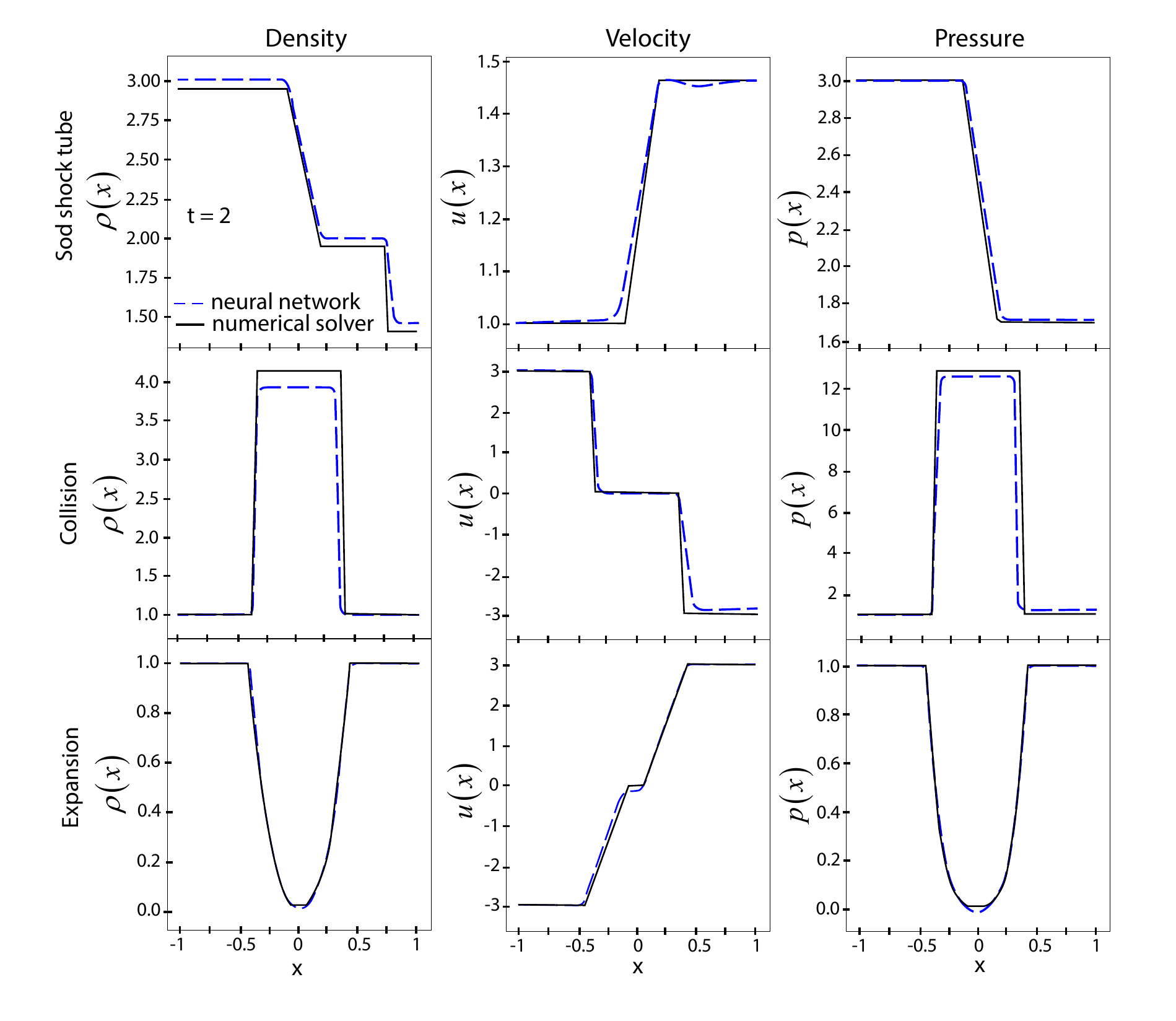}
\caption{\label{test}
Comparison of the DNN performance (red solid lines) with the analytical solutions of the numerical solver (blue dashed lines) for the following Riemann Problem Test cases: (top) Sod Shock Tube, (middle) collision, and (bottom) expansion at $t = 2$. Here $\rho(x)$, $u(x)$, $p(x)$ are the dimensionless density, velocity and pressure dependencies correspondingly.
}  
\end{figure*}

In addition to the training set accuracy, we also subjected the network to the Riemann Problem~\cite{menikoff1989riemann}, which is a classical test problem for evaluating the accuracy and stability of a hydro solver.
For this purpose the training set has been extended to incorporate a family of profiles that are similar to the Riemann Problem's initial conditions.
The comparison of the DNN generated profiles with an accurate numerical solution corresponding to the classical cases of the Sod shock tube Riemann case~\cite{monaghan1983shock}, a collision as well as expansion scenario is depicted in top, middle, and bottom panels of Fig.~\ref{test}, respectively.
While a few percent level deviations from the numerical solution are visible, the network captures accurately all the qualitative features of the Riemann Problem solutions.

\begin{figure}[h!]
\includegraphics[width=0.49\textwidth]{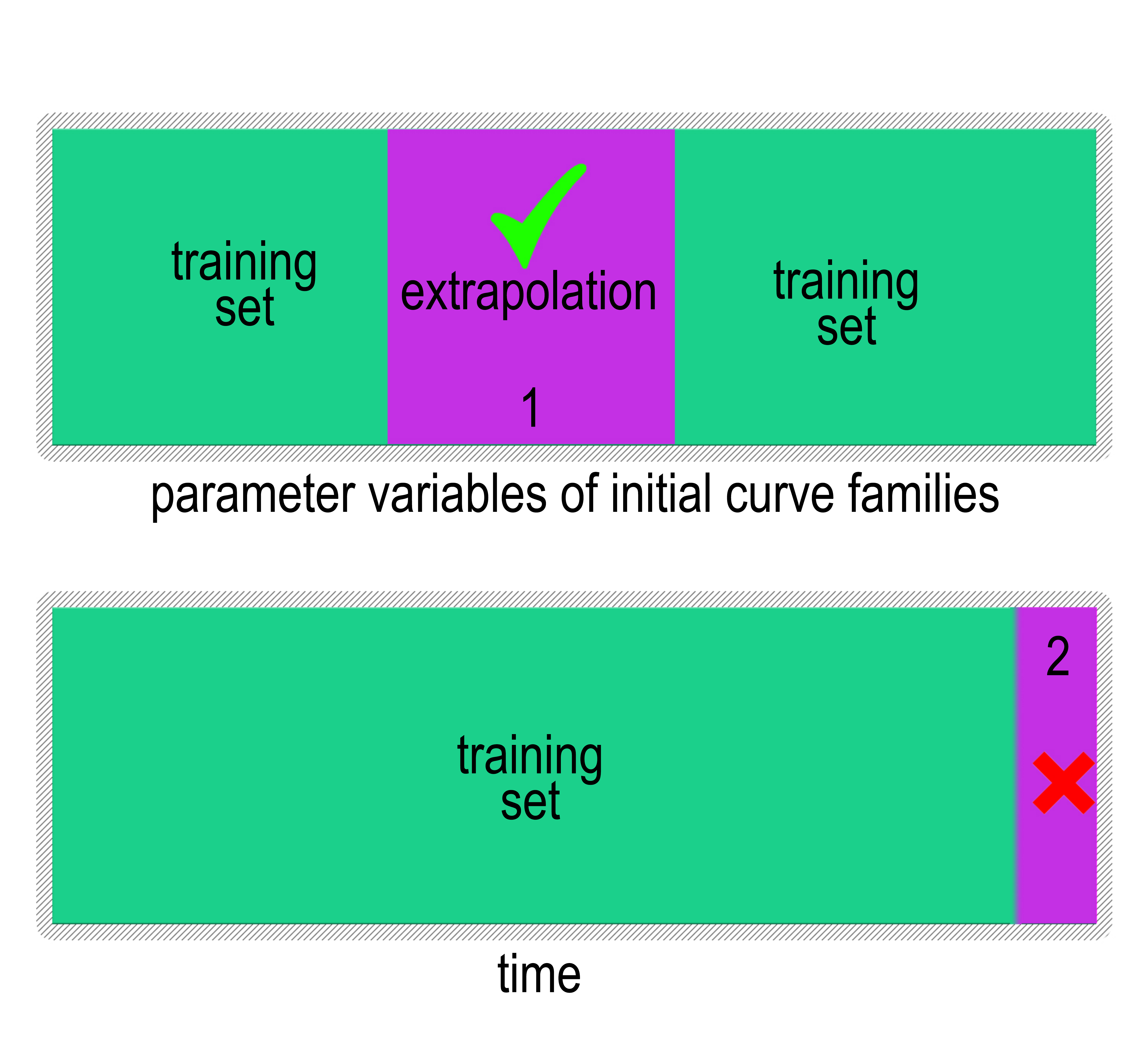}
\caption{\label{extrapolation_reg}
Regions in the parameter space of the initial hydro profiles and regions along the spatial and time axes in which the network's extrapolation ability was studied. The regions in which the network demonstrated positive performance are marked with the check mark. The regions, where the results are negative are marked with the red cross. The performance of the network of the present structure is still acceptable just outside the training set. But shortly after the end of the training set instabilities start to emerge. This behavior is demonstrated with the corresponding gradient. 
}
\end{figure}

\begin{figure}[h!]
\includegraphics[width=0.4\textwidth]{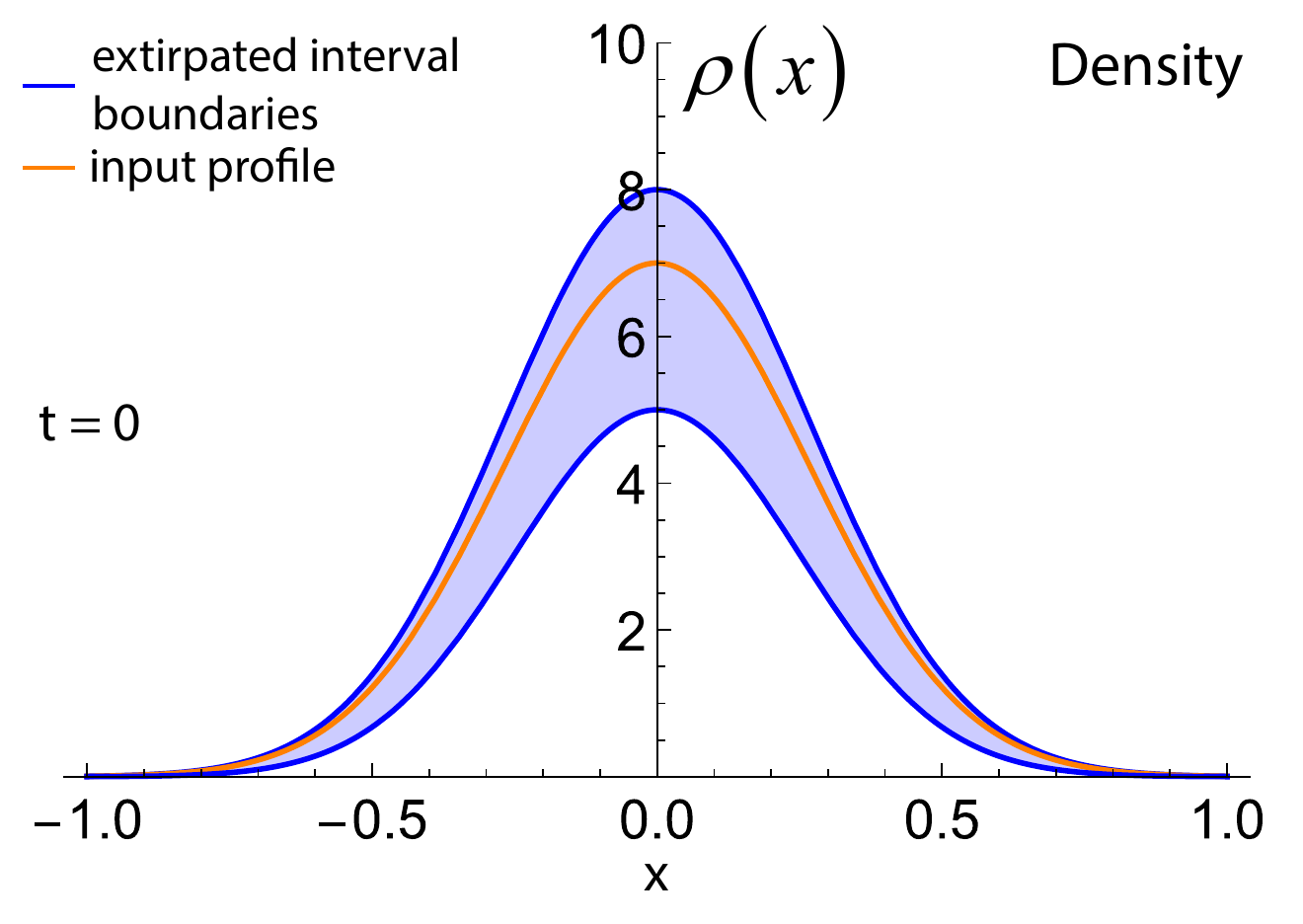}
\includegraphics[width=0.4\textwidth]{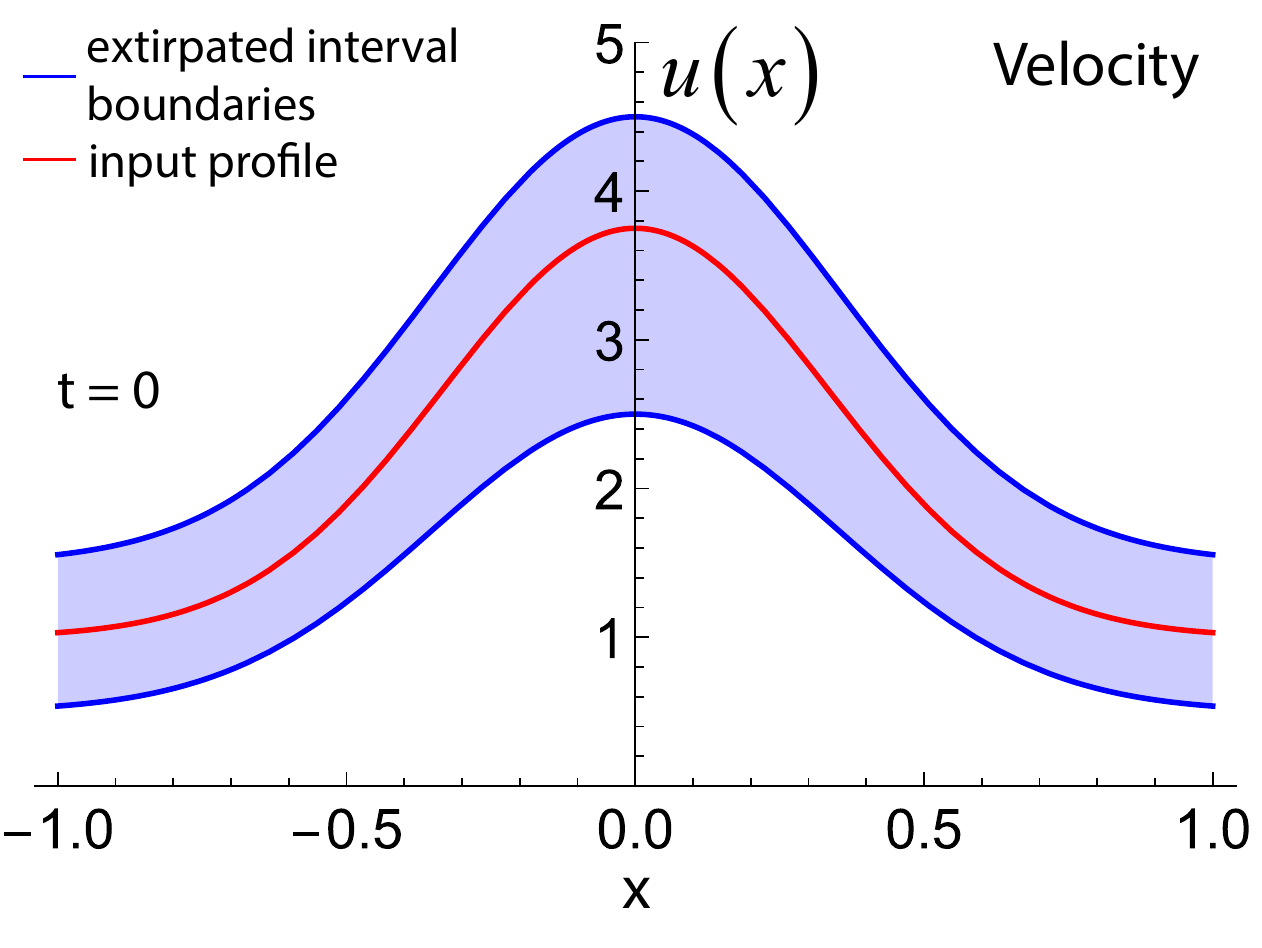}
\caption{\label{initial}
Initial density and pressure profiles, which are contained within the interval, that was removed from the complete test set in order to test the network's capacity to interpolate. The curves from the extirpated interval are used as the input for the network with the diminished training set.
}
\end{figure}
\begin{figure}[h!]
\includegraphics[width=0.49\textwidth]{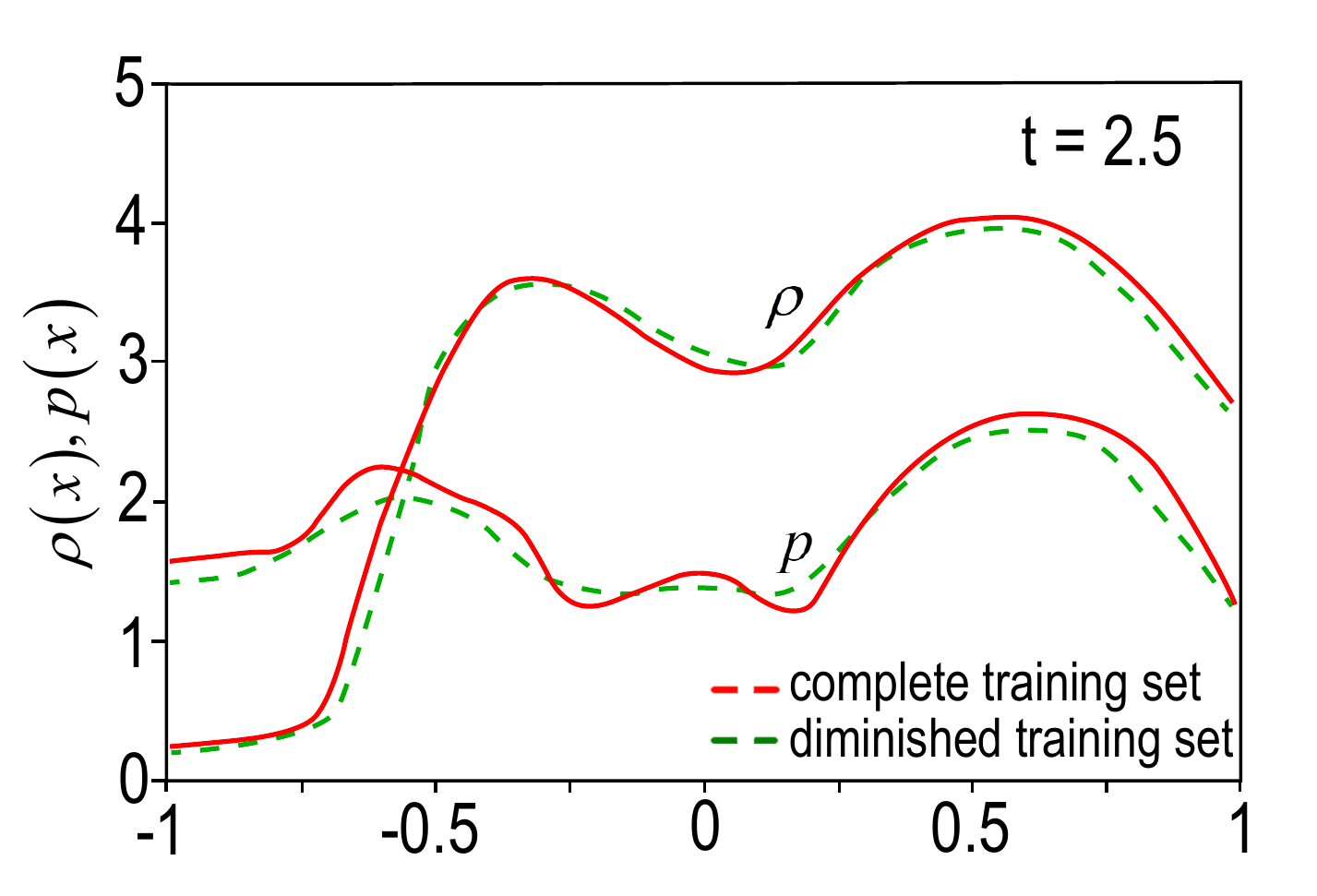}
\caption{\label{performance}
Comparison of the performance of the two neural networks  at $t = 2.5$ -- the one trained on the complete training set, containing the initial density and pressure profiles shown in Fig.~\ref{initial} (red solid lines) and another trained on the diminished training set, which does not contain the aforementioned subset of curves (green dashed lines). The initial density and pressure profiles at $t = 0$ used for this test case are shown in Fig.~\ref{initial}.
}
\end{figure}
\begin{figure*}
\includegraphics[width=0.86\textwidth]{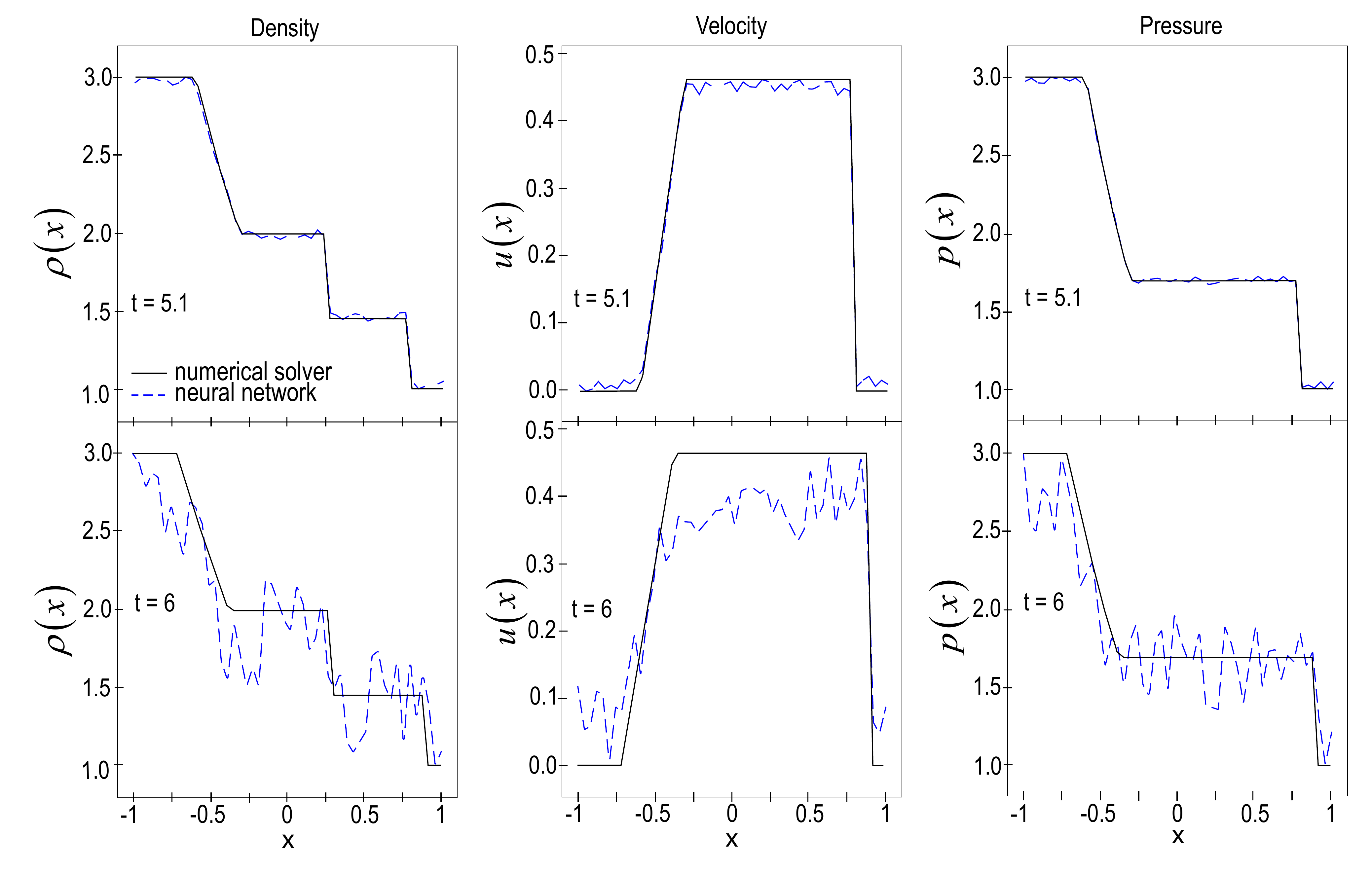}
\caption{\label{dvp3}
Comparison of the DNN performance (red solid lines) with the analytical solutions of the numerical solver (blue dashed lines) at time $t = 5.1$ (top) and $t = 6$ (bottom) which is beyond the training region $t \in [0,5]$. 
The performance of the DNN with a variable time-shift parameter  deteriorates right beyond the training interval. 
}
\end{figure*}

\section{DNN Performance outside the training set}
\label{sec-outside}

We have shown that the network is able to approximate the solution with high accuracy when the solution is within the boundaries of the training interval.
An important question now is: how good are the network capabilities to extrapolate outside the training set?
Here we study three different aspects of the network's extrapolation ability:
\begin{enumerate}
    \item The performance on hydro profiles that are not included in the training set
    \item The ability to extrapolate to time intervals not included in the training set
\end{enumerate}
These two cases are schematically depicted in the top and bottom panels of Fig.~\ref{extrapolation_reg}.

First, we explore the extrapolatory ability of the network to cover profiles not included in the training~(region "1" in Fig.~\ref{extrapolation_reg}).
This is achieved by excluding a certain subset of initial $\rho,u,p$ curves~[Eqs.~(\ref{f1})-(\ref{f3})] from the training set, which correspond to a particular interval of the parameters ${a_i}$, ${b_i}$, ${c_i}$.
Namely, we omit density profiles corresponding to parameter intervals $a\in[5,8]$, $b\in[-7,-8]$ for the $f_1$ family of initial curves and velocity profiles corresponding to $a\in[2,3]$, $b\in[-4,-5]$, $c\in[0.5,1.5]$ for the same $f_1$ family of initial curves.
We then analyze the performance of such a network on an input profile that was omitted from the training set.
This procedure is illustrated in Fig.~\ref{initial}.
To assess the extrapolation performance of the network, we compare its output to the one produced by a network based on the full training set.
The comparison, presented in Fig.~\ref{performance}, indicates that the network performance is virtually unaffected by excluding a broad set of profiles from the training data.\footnote{Note, that this interpolation test differs from the classical “test-set” trial of the network performance, when just some random samples are excluded from the training set and are moved to the test set. Given the vast initial training set of $2.4~10^5$ curves,
even with multiple of the curves being 
excluded,
there will still remain plenty of quite close curves.
}.
Thus, the network demonstrates reasonable ability to extrapolate from the different initial profiles.

Next, we study the extrapolation ability along the time axis.
Namely, we analyze the network output for time moments exceeding the training interval of $t \in [0,5]$~ (region "2" in Fig.~\ref{extrapolation_reg}).
We take two values: (i) $t = 5.1$ which is slightly outside the training interval and (ii) $t = 6$ which is further away.
The results for these two cases are depicted in top and bottom panels of Fig.~\ref{dvp3}, respectively.
The network performance is still relatively decent in the first case just outside the training interval, although the developing instabilities are visible.
For $t = 6$ the network output is quite off from the reference solution, at best capturing only the general qualitative features of the solution.
This indicates that the extrapolation ability of the present network with a variable time step is weak.
For this reason we explore a different treatment of the time step.
It is also possible to explore other scenarios of interpolation. For instance, when a certain period of time-evolution of the system was omitted and must be reconstructed.

\section{Fixing the time step} 
\label{sec-fix-steps}

 \begin{figure*}
\includegraphics[width=0.86\textwidth]{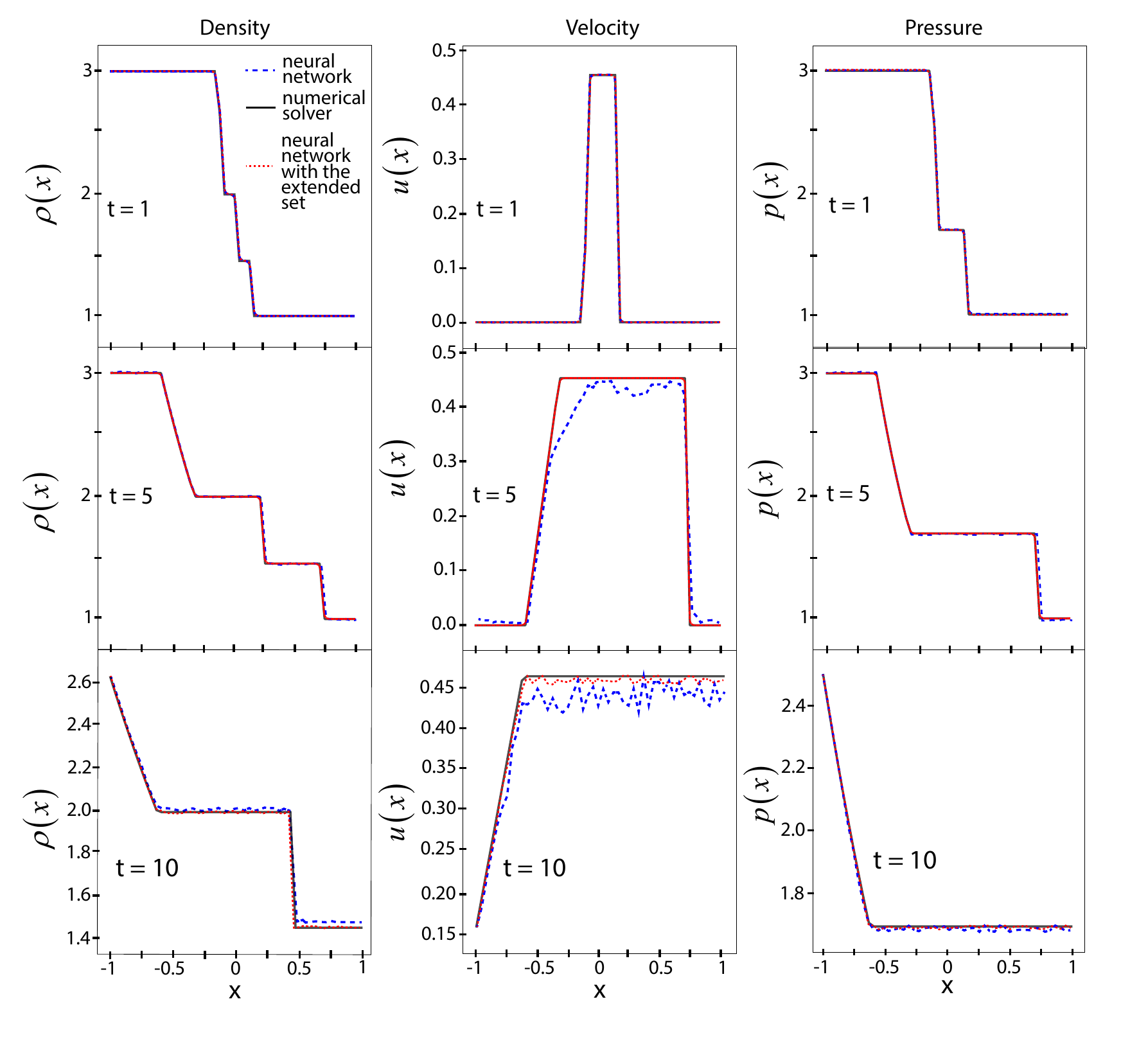}
\caption{\label{dvp}
Comparison of the altered DNN performance trained on the original training set (red solid lines) and extended training set (blue solid lines) with the analytical solutions of the numerical solver (green dashed lines)
for the Sod shock tube problem at time parameter values $t = \delta t \equiv 1$ (top), $t = 5$ (middle), and $t = 10$ (bottom).
} 
\end{figure*}

As shown in the previous section, the DNN performance along the time axis becomes unsatisfactory as one goes beyond the training range.
This is most likely a reflection of an unbalanced structure of neurons in the network used in Secs.~\ref{sec-meth} and \ref{sec-outside}.
Whereas $50 \times 3$ parameters were used for the spatial variable, only a single neuron in the input layer was dedicated to the time variable. 
This one neuron alone defines the whole shape of the output curves, as each different time-shift $\delta t$ corresponds to different $\rho, u, p $ mappings.
The output $\rho, u, p $ profiles strongly depend on the time-shift $\delta t$ parameter.
This leads to an
asymmetry in the distribution of weights and biases, as this single, time shift dedicated, neuron in the input layer 
is significantly
more important than any other. 
The asymmetry in the “significance” leads to non-uniform distribution of weights and biases encountered by the network in the input layer, consequently affecting its ability to extrapolate.

To mitigate the situation we modify the network structure in this section.
We do not specify a variable time-shift and remove the corresponding neuron from the DNN.
Instead we specify a fixed 
time shift between the initial and the final states:
$\rho \left( {x,t} \right),u\left( {x,t} \right),v\left( {x,t} \right) \to \rho \left( {x,t + \delta t} \right),u\left( {x,t + \delta t} \right),v\left( {x,t + \delta t} \right)$. 
In this way we eliminate the non-uniformity in the distribution of values in the input layer.
This also reduces the required informational capacity, as the network only needs to learn the mapping for a single fixed time step rather than for a multitude of steps.

The training of the new network is performed similarly to the previous one.
The training data consists of the $\rho, u, p $ profiles mapped between the initial time moment $t_0$ and a subsequent time moment $t + \delta t$ with $\delta t = 1$.
These profiles are combined into training pairs 
\eq{
\{\rho_0(x),u_0(x),p_0(x)|\rho_{0.1}(x),u_{0.1}(x),p_{0.1}(x)\}
}
and the training is performed.

The updated network allows to perform mapping $f\left( {x,{t_0}} \right) \to f\left( {x,{t_0} + \delta T} \right)$ across large time steps $\delta T$, provided that it can be presented as a multiple of $\delta t$, i.e. $\delta T = n\delta t$, $n \in N$.
This is achieved by successively applying the network $n$ times, by using the output from the previous step as an input to the next one.
To test the performance of the new network along the time axis we will subject it to the Sod shock tube test.
More specifically, we study the performance in a time interval $t  \in  [0,10]$ by consequently performing 10 steps in increments of $\delta t = 1$ as described above.

The performance of the altered network is presented in Fig.~\ref{dvp} (solid red lines) and compared to the numerical solution~(dashed blue lines).
The top panel corresponds to $t = 1$, this short time interval coincides with the training interval, thus, it illustrates the interpolating capability of the network.
The network reproduces the numerical solution with high accuracy, thus its interpolating ability is good.
The middle and bottom panels of Fig.~\ref{dvp} correspond to $t = 5$ and $t = 10$, respectively.
These intervals are outside the training range, thus, the extrapolation capability of the network is probed.
The updated network clearly 
shows much better capacity to 
extrapolate
in comparison with the previous method employed in Secs.~\ref{sec-meth} and \ref{sec-outside}.
However, already for $t = 5$ one can see deviations of the network generated profiles from the numerical solver, especially for the velocity profile.
Thus, further improvements are required.

 \section{Extending the training set}
 \label{sec-extend-training}
 
The DNN performance can be further improved by modifications to the training procedure.
Here we explore two modifications: (i) an extended set of training profiles and (ii) training the network across several time steps.
The first one addresses a possible gap in the set of profiles that the network is able to map accurately while the second modification is designed to improve the network extrapolation performance across large time periods.

\subsection{Extending the training set with harmonic functions}
\label{sec-extend-1}
 
The results of the previous section show considerable error accumulation as the number of time steps is increased, in particular in the velocity profile.
We surmise that the error accumulation energy due to an incomplete range of velocity profiles covered in the training set, namely that the $f_{1,2,3}$ curve families used in training are not close enough to accurately capture the features of the velocity profile in the Sod shock tube test.
Therefore, here we extend the training set with an additional family of curves, corresponds to a linear combination of harmonic functions:
\eq{\label{f4}
f_4 = \sum\limits_{i = 1}^n {{a_i}\sin \left( {{b_i}x + {c_i}} \right)}.
}
Here $a_i$, $b_i$, $c_i$ are random parameters chosen uniformly
for the following ranges:
\eq{\nonumber
{a_i}\in \left[ { 0,10} \right],~{b_i} \in \left[ { - 15,15 } \right],~{c_i}\in \left[ { - \pi ,\pi } \right],~{n}=1,...,10 .}

\begin{figure}[h!]
\includegraphics[width=.49\textwidth]{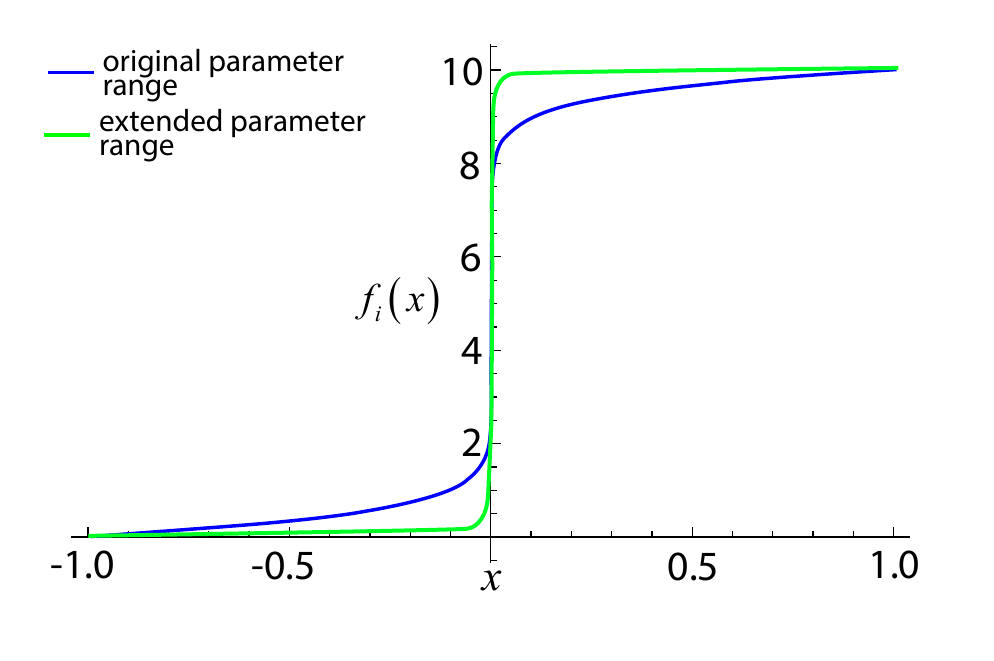}
\caption{\label{rangeextension}
An illustration of how the extended parameter range allows to better approximate discontinuous profiles.
Here the blue and green curves show the best approximation achievable by, respectively, the original and extended parameter sets for a step function $f(x) = 10 \, \operatorname{\theta}(x)$.
}
\end{figure}

With this extension the training set allows a significantly more accurate coverage of discontinuous profile functions, as illustrated in Fig.~\ref{rangeextension} for the case of a step function and similar discontinuous profiles. 
We generated $2.5\times 10^5$ samples for training and $8\times 10^4$ for validation of the modified network.
For this network structure without the time-shift parameter
in the input layer with the periodic boundary conditions
within $24$ training epochs it was possible to obtain an accuracy level of $92\%$.

The performance of the modified DNN for the Sod shock tube test is shown in Fig.~\ref{dvp} by the solid green lines.
One sees that the extension of the training set for the velocity significantly improves the performance of the network.
The improved performance is particularly
evident at
$t=10$, which corresponds to consecutive 10 network iterations.

\subsection{Training over multiple time steps}

\begin{figure}
\includegraphics[width=.49\textwidth]{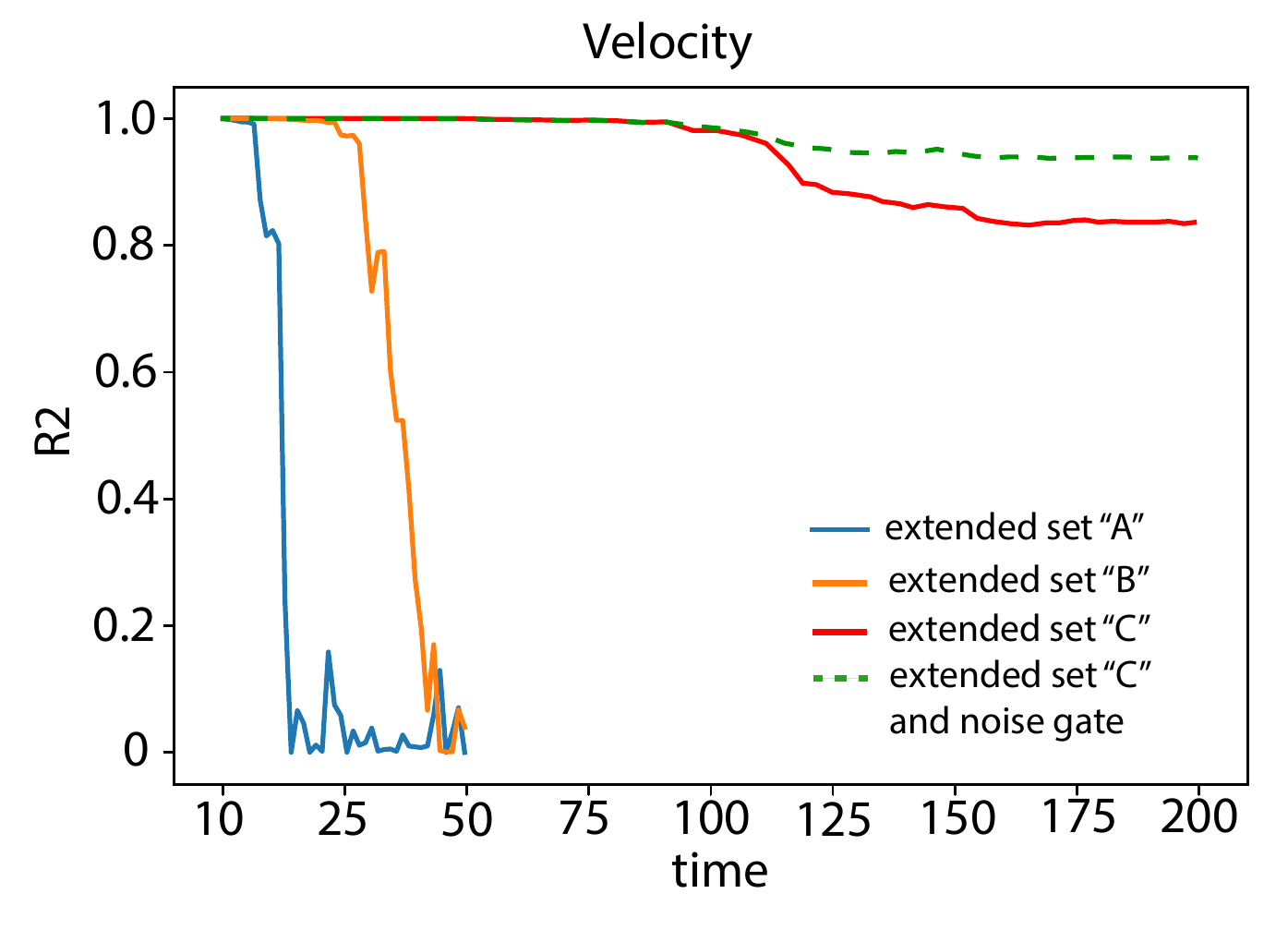}
\caption{\label{r2_metric}
The dependence of the $R^2$ metric between the solver and network performance on the time (number of network iterations) for the velocity profile for different training sets used. Set "C" includes 6 consequent generations of curves computed at the corresponding time values from ($t_0$) to ($t_0+6*\delta t$). Set "B" includes only one additional time step($t_0+\delta t$). Set "A" presents only the modified network structure with no time-shift parameter in the input layer. Additionally the effect of the noise gate filter application is shown. Note that the alterations in the structure of training sets alter the content of the set but not the total number of curves present in it.
}
\end{figure}

\begin{figure}
\includegraphics[width=.4\textwidth]{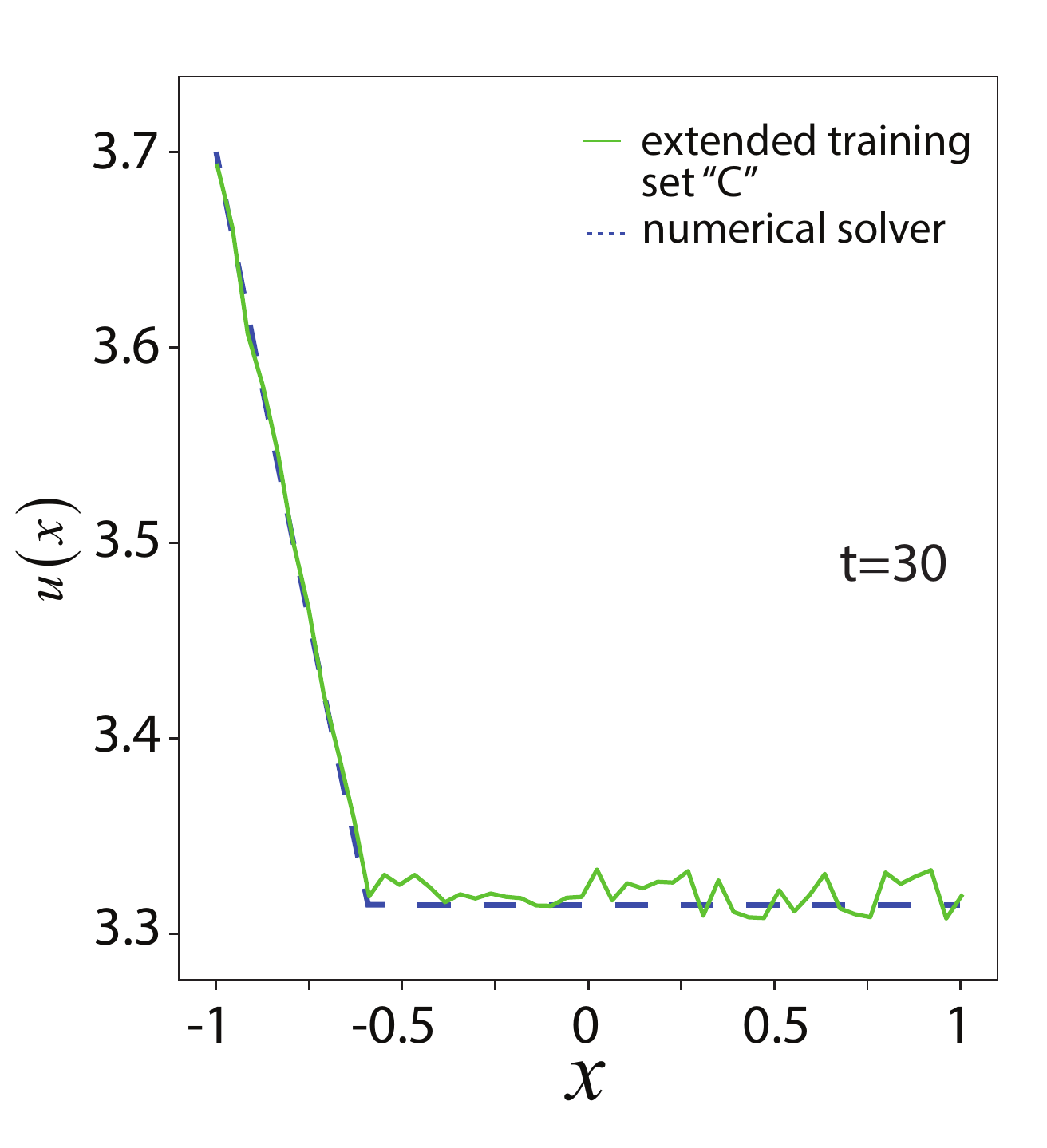}
\caption{\label{further_set_extension}
The effect of further extension of the training set at t=30 (after 30 network iterations) on the velocity dependency.}
\end{figure}

Another extension of the training set is achieved by utilizing multiple time steps in the training procedure.
This is achieved in the following way.
We take the initial profiles from the extended data-set $A$ described in Sec.~\ref{sec-extend-1}.
Then, for each curve the numerical solver is run up to a time moment $t = 6$.
We then take the profiles at time moments $t = 0$, 1, 2, 3, 4, 5, 6, and include them into the training set.
In this way the new extended training set ($C$) contains not only the set of profiles $f_1,...,f_4$ and their mapping to a subsequent $t+\delta t$ time moment, but also that for the profiles obtained from each of the steps from one ($t+\delta t$) to six ($t+6*\delta t$). 
The training utilizing this extended set is then done from scratch.
It is observed that the extended training set $C$ leads to a significantly improved performance of the network across long time intervals.
This is illustrated in Fig.~\ref{r2_metric}, which shows the behavior of the $R^2$ error metric for the different studied training sets across a long time interval extending to $t = 200$.
The error increases substantially after a certain number of network iterations performed. 
This behavior is attributed to encountering new curve configurations, that are not anymore covered accurately by the standard training set ($A$). 
The training set $B$ incorporates a single additional time step $t+\delta t$.
This allows to preserve a satisfactory network performance to larger times, however the accuracy drop substantially at $t ~ 30-50$.
The full extended training set "C", on the other hand, preserves a relatively constant level of the $R^2$ error metric across long time intervals.
Figure~\ref{further_set_extension} compares the velocity profile output of the network based on training set $C$ at $t = 30$ with the numerical solution.

A substantial performance increase due to the extension of the training set with profiles from multiple time steps is evident.
This is due to the fact, that the training set $A$,
which did not contain training profiles after multiple time steps, 
remained incomplete,
i.e. the set of trained configurations was not sufficient for
the network to perform the mapping at long time intervals.
The result effectively underlines that the particular choice of the initial training set appears to be one of the most important factors influencing the capability of the method.
This choice will generally depend on multiple other factors, including the   geometry of the system, equation of state and boundary conditions. 
In this work we use the periodic boundary conditions, if these are changed to other boundary conditions such as open or hard-walls, the network should be re-trained.
It should also be emphasized that here we trained the network for a fixed value of the time shift, thus the network is only capable of performing mappings in the units of this fixed time shift.
If a different value of the time shift is desired, this will also require the re-training of the network.

\section{noise gate filter}
\label{sec:noisegate}

As discussed in the previous section, the performance of the network drops after a large number of time steps even in the case of the most complete training set "C".
This is reflected by the appearance of noise in the hydrodynamic fields~(Fig.~\ref{further_set_extension}).
Although the level of noise remains fairly stable, it does notably affect the performance of the method, the $R^2$ accuracy metric reaching about $0.83$ at large times, as seen in~Fig.~\ref{r2_metric}.
Here we explore a possibility to alleviate this issue by implementing a noise gate filter.
The benefit of knowing the numerical solution along with the prediction of the neural network allows one to train the separate convolutional autoencoder structure, which performs the role of a noise gate filter. 

This noise gate filter is incorporated in the following way.
First, we generate multiple clean training curves belonging to the aforementioned 6 iterations of evolution from $t_0$ to $t_0+i*\delta t$, where      ${i} \in \left[ { 0,6} \right]$. 
Then, we prepare the training input-output pairs by adding (i) random Gaussian noise of various  types, (ii) uniformly distributed noise and (iii) noiseless samples.
The latter ensures that the autoencoder learns not to filter the already clean samples. 
$10^5$ training pairs are prepared in this way, then this autoencoder structure is applied to the output obtained from the neural network trained on the extended set "C" in the previous section. 
The resulting effect of the autoencoder on the performance of the network trained on the extended set "C" in the previous section is depicted in Fig.~\ref{r2_metric} by the dashed green line.
The application of the noise gate filter considerably improves the network accuracy at large times $t > 100$.

\begin{figure*}
\includegraphics[width=0.8\textwidth]{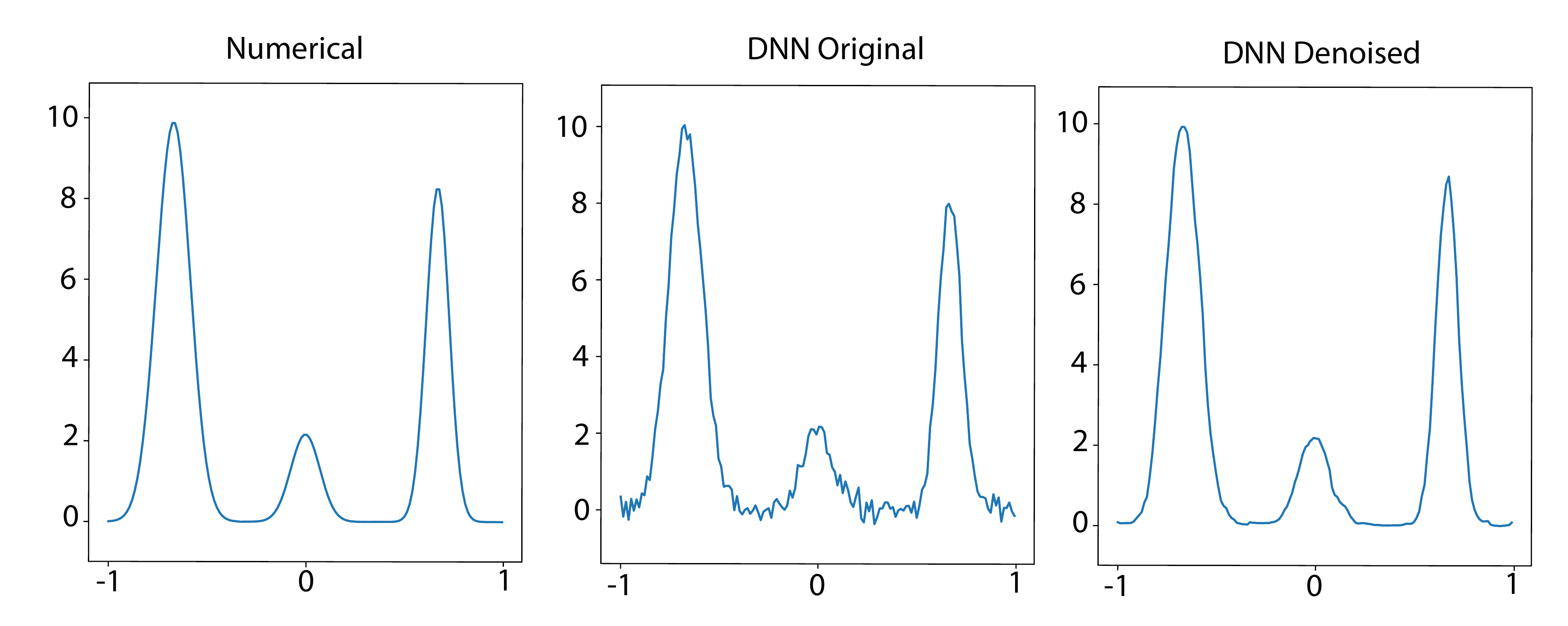}
\caption{\label{denoising_process}
The effect of the application of the noise gate layer on the velocity profile. The velocity profile containing noise is presented in the middle panel. The result of the noise - gate layer application is presented in the right panel and the conventional reference solution from the numerical scheme is presented at the left panel. 
}
\end{figure*}

The main benefit of the noise gate filter is to smoothen the hydro profiles.
This is illustrated in Fig.~\ref{denoising_process}, which depicts a typical case of a network-produced velocity profile with and without the application of the noise gate filter compared to the reference numerical solution.
While the output of the regular network is able to capture the overall numerical profile, it does exhibit characteristic noise pattern.
The noise gate filter is able to largely remove this noise and resemble the reference profile even more accurately.
The application of the noise gate thus allows to substantially increase the accuracy of the neural network based model.

\section{performance}
\label{sec:performance}

The main purpose of applying the DNN to fluid dynamics is to improve the speed of the associated computations over the conventional methods while preserving an acceptable level of accuracy. 
This can be quantified by a speed-up factor $s$.
Its value will depend on the particular task under consideration.
Let us consider the following generic problem: the fluid dynamical simulation is run over a time period $\delta t_e$, with an intermediate output at $n_{\rm int}$ time steps.
For instance, such a setup can occur in a case of real-time simulation of the flow for visualization purposes, where at least 24 frames per second are required to render.
The maximum speed of the conventional method can be achieved by maximizing the value of a single time-step, which will minimize the total number of steps needed to be performed.
This maximum value, $\delta t_{\rm CFL}$, is constrained by the Courant–Friedrichs–Lewy~(CFL) condition.
The neural network, on the other hand, is not constrained by the CFL condition.
Instead, the time step per single DNN iteration is $\delta t_e / n_{\rm int}$, reflecting the fact that one needs to output $n_{\rm int}$ intermediate time steps.
The speed-up factor thus reads
\eq{
s = \frac{{\delta {t_{\rm{e}}} / n_{{\rm{int}}}}}{{\delta {t_{{\rm{CFL}}}}}} \, \frac{{{\tau _{\rm num}}}}{{{\tau _{\rm net}}}}~.
}
Here $\tau_{\rm num}$ is the time duration of a single computational step that computes the evolution from $t^*$ to $t^*+\delta t_{\rm CFL}$ and $\tau_{\rm net}$ is the time duration of a single computational step of neural network application.

The performance increase due to the neural network comes from two sources.
First, the neural network application is not constrained by the CFL criterion, thus $\frac{{\delta {t_{\rm{e}}} / n_{{\rm{int}}}}}{{\delta {t_{{\rm{CFL}}}}}} > 1$ is possible.
Second, the neural network can perform a single time step considerably faster than the conventional methods.
For the applications considered in the present work, we have ${\tau_{\rm net}}/{\tau_{\rm num}} \sim 10^2$.
The values of $\frac{{\delta {t_{\rm{e}}} / n_{{\rm{int}}}}}{{\delta {t_{{\rm{CFL}}}}}}$ were observed to vary in range $15-50$, depending on the exact network configuration employed.
The speedup factors listed here correspond to both the numerical method and the network running on CPU.
Additional speed-up up to a few orders magnitude can be obtained by employing hardware tailored for DNN-related computations, such as NVIDIA tensor core GPUs.

\section{summary}
\label{summary}

We investigated the possibility of a deep neural network based simulation of one-dimensional fluid dynamics.
The DNN performance vastly depends on the network structure and training set, an acceptable accuracy can be obtained with an appropriate choice.
For the network structure, a network with 3 hidden layers and one dropout layer was found to be a decent choice, yielding a 90\% accuracy on the training set that consists of a vast amount of fluid dynamical profiles.
Regarding the network's ability to extrapolate along the time axis, it shows a limited ability when the time step is varied and part of the network input.
On the other hand, a decent performance is achieved when the time step is fixed and the network is applied successively by using the output from the previous time step as input to the next step.
The performance is further optimized by extending the training set with profiles obtained numerically after multiple time steps.
We also find that the noise in the profiles output by the network at future times can be to a large extend eliminated by the application of a noise gate filter.

The time step of the DNN simulations is not restricted by the Courant criterion, which is a major source of the speed-up of DNN computations over the conventional numerical methods.
Also, the time spent per single time step in the DNN is observed to be about two orders of magnitude smaller than that for the conventional computations.
This performance can be further improved substantially by utilizing dedicated hardware tailored for DNN applications, such as modern GPU's with tensor cores. 
While the presented studies indicate that a careful choice of the network structure and the training set is not trivial and crucial to achieve an acceptable performance, it is reasonable to expect even greater benefits of the DNN in higher dimensions, where the conventional methods become very time consuming.

\section*{Acknowledgments}

K.T. and R.P. express gratitude to the Stiftung Polytechnische Gesellschaft for supporting the research project.
K.Z. and J.S. thank the AI grant at  FIAS by SAMSON AG and the BMBF funding through the ErUM-Data project.
R.P. acknowledges the support of the Target Program of Fundamental Research of the Department of Physics and Astronomy of the National Academy of Sciences of Ukraine
(N 0120U100857).
V.V. acknowledges the support through the
Feodor Lynen program of the Alexander von Humboldt
foundation and the U.S. Department of Energy, 
Office of Science, Office of Nuclear Physics, under contract number 
DE-AC02-05CH11231231. H.St. acknowledges the Walter Greiner Gesellschaft zur F\"orderung der physikalischen Grundlagenforschung e.V. through the Judah M. Eisenberg Laureatus Chair at Goethe Universit\"at Frankfurt am Main. K.Z. also thank NVIDIA Corporation with the generous donation of NVIDIA GPU cards.

\appendix

\section{Network description}
\label{app-a}

\begin{figure}[h!]
\includegraphics[width=.33\textwidth]{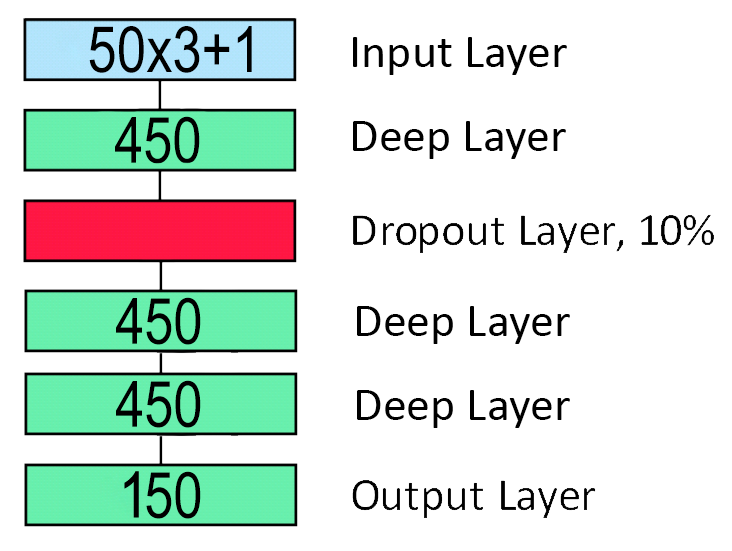}
\caption{\label{structure}
The DNN structure used in the present work.
}

\end{figure}

We consider two slightly different network structures: with and without a variable time-shift parameter.
The general network structure for both cases is depicted in Fig.~\ref{structure}.
The input layer contains three function profiles, $\rho(x)$, $u(x)$, and $p(x)$.
Each of the profiles is discretized into 50 equidistant points and contributes 50 corresponding neurons to the input layer.
In addition, the network with a variable time shift contains an additional neuron in the input layer specifying the value of the time shift parameter.
Thus, in total there are 151~(150) neurons in the input layer for the network with variable~(fixed) time shift.
The network consists of three hidden layers, each containing 3*150 neurons, one dropout layer (rate 10\%), and the output later with 150 neurons corresponding to the $\rho(x)$, $u(x)$, and $p(x)$ at the next time step.

The training can be performed utilizing any desired set of profiles.
We used $2.4\times10^5$ and $8\times10^4$ randomly generated initial input profiles for $\rho,u,p$ for, respectively, training and validation.
The generation proceeds in the following way: we first arbitrary select the curve family from one of the three curve families, namely, $a~{\rm exp}(b x^2)$, $a~{\rm sin}(b x + c)^2$, $a|x^{\frac{1}{b}}|sgn(x)+c$. 
These curve families are chosen to emulate generic possible initial distributions of $\rho,u,p$. Having selected the particular function family we then randomly select $a,b,c$ coefficients within the corresponding ranges for each family of functions. In this way density and velocity profiles are randomly generated and the resulting $p$ curve is computed in accordance with the properties of the regarded system to assure consistency with the equation of state.

Having generated a substantial training and validation sets of input profiles we have to supplement input curves with the output curves. 
This procedure is different for two different network structures we employ in the present paper. The network a variable time-shift parameter requires the training process to include many mappings 
$\rho(x,t_0),u(x,t_0),p(x,t_0)\rightarrow \rho(x,t_0+\delta t),u(x,t_0+\delta t),p(x,t_0+\delta t)$ as the training pairs  for various $\delta t$ values within the chosen $\delta t \in [0,5]$ interval.
A particular time-shift value is uniformly selected within the chosen training time-interval. 
Then, the generated initial profiles are supplied into the numerical hydro-solver which calculates the evolution of the input curves up to the time value specified by the time-shift parameter. 
The output curves calculated by the solver are used to supplement the input profiles with the output profiles and form the input-output training pairs,
\eq{\nonumber
&\{\rho(x,t_0),~u(x,t_0),~p(x,t_0),~t^*\},\\\nonumber
&\{\rho(x,t^*),~u(x,t^*),~p(x,t^*)\}
}
Such procedure is performed for all training and validation input profiles. 

The network with the fixed time step, on the other hand, does not contain the time-shift parameter in the input layer. 
Thus, the training procedure is slightly different. 
The network always performs the mapping for a fixed, constant value of the time shift parameter. 
In this case the generated initial profiles are supplied into the numerical hydro-solver which calculates the evolution of the input curves always up to this fixed time-shift value. 
This procedure is also performed of all training and validation input profiles. 

\section{The numerical solver description}
\label{app-b}

The numerical solver is used as a reference solution employed for training and validation.
We use a traditional Riemann solver described e.g. in~\cite{toro2013riemann}~(using the implementation from~\cite{zingale2014introduction}).
The spatial domain is broken into the discretized units. 
The time-wise discretization is performed in accordance with the Euler scheme with the explicit time integration. 
The pressure field is assigned to the centers of the computational units, whereas the velocity field is computed at the edges. 
The numerical simulation is performed in the real space formulation and the advection approach is based on the usage of the finite-difference scheme for both convection and diffusion.

\bibliography{main.bib}

\end{document}